\documentclass[twocolumn,showpacs,preprintnumbers,amsmath,amssymb]{revtex4}

\usepackage{graphicx}
\usepackage{dcolumn}
\usepackage{bm}

\begin{document}

\preprint{}

\title{Steep sharp-crested gravity waves on deep water}

\author{Vasyl' Lukomsky}
\email{lukom@iop.kiev.ua}
\author{Ivan Gandzha}
\author{Dmytro Lukomsky}
\affiliation{Department of Theoretical Physics, Institute of
Physics, Prospect Nauky 46, Kyiv 03028, Ukraine}

\date{November 12, 2001}

\begin{abstract}
A new type of steady steep two-dimensional irrotational symmetric
periodic gravity waves on inviscid incompressible fluid of
infinite depth is revealed. We demonstrate that these waves have
sharper crests in comparison with the Stokes waves of the same
wavelength and steepness. The speed of a fluid particle at the
crest of new waves is greater than their phase speed.

\end{abstract}


\maketitle

A proper understanding of various wave phenomena on the ocean
surface, such as modulation effects and instabilities of large
amplitude wave trains \cite{Schwartz:review}, \cite{McLean},
formation of solitary \cite{Camassa}, freak \cite{Osborne:freak},
and breaking waves \cite{Peregrine:review,Saffman-Yuen}, requires
knowledge of a form and dynamics of steep water waves. For the
first time surface waves of finite amplitude were considered by
\citet{Stokes}. Stokes conjectured that such waves must have a
maximal amplitude (the limiting wave) and suggested that a free
surface of the limiting wave near the crest forms a sharp corner
with the $120^{\circ}$ internal angle (the Stokes corner flow). A
strict mathematical proof of the existence of small amplitude
Stokes waves was given by Nekrasov. \citet{Toland} proved that
Nekrasov's equation has a limiting solution describing a
progressive periodic wave train which is such that the flow speed
at the crest equals to the train phase speed, in a reference frame
where fluid is motionless at infinite depth. \citet{Long:al-h1977}
constructed asymptotic expansions for waves close to the
$120^{\circ}$-cusped wave (almost highest waves) and showed that
the wave profile oscillates infinitely as the limiting wave is
approached. Later, in \cite{Long:al-h1994-2}, the crest of a
steep, irrotational gravity wave was theoretically shown to be
unstable.

The purpose of the present work is to give evidence that a second
branch of two-dimensional irrotational symmetric periodic gravity
waves of permanent form exists besides the Stokes waves of the
same wavelength. The original motivation is as follows: the
Bernoulli equation is quadratic in velocity and admits two values
of the particle speed at the crest. The first one corresponds to
the Stokes branch of symmetric waves for which the particle speed
at the crest is smaller than the wave phase speed. The opposite
inequality takes place for the second branch which might
correspond to a new type of waves. In the second part of the
Letter, we prove this numerically by using two different methods.

Consider a symmetric two-dimensional periodic train of waves which
propagates without changing a form from left to right along the
$x$-axis with the constant speed $c$ relative to the motionless fluid at
infinite depth. The set of equations governing steady potential
gravity waves on a surface of irrotational, inviscid,
incompressible fluid is
\begin{eqnarray}
\label{eq:Lapl}
 \mathit{\Phi}_{xx}+\mathit{\Phi}_{yy}=0, && -\infty<y<\eta(\theta);
\\ \label{eq:Dyn}
 \left(c-\mathit{\Phi}_{x}\right)^2+\mathit{\Phi}_y^2+2\eta=c^2,
 && y=\eta(\theta);
\\ \label{eq:Kin}
 \left(c-\mathit{\Phi}_{x}\right)\eta_{x}+\mathit{\Phi}_y=0, && y=\eta(\theta);
\\ \label{eq:Inf}
 \mathit{\Phi}_x=0,~\mathit{\Phi}_y=0,~y=-\infty; &&
 \theta=x-ct.
\end{eqnarray}
where $\mathit{\Phi}(\theta,~y)$ is the velocity potential,
$\eta(\theta)$ is the elevation of a free surface, and $y$ is the
upward vertical axis such that $y=0$ is the still water level. We
have chosen the units of time and length such that the
acceleration due to gravity and wavenumber are equal to unity.

As it follows from the Bernoulli equation (\ref{eq:Dyn}), a
solution may be not single-valued in the vicinity of the limiting
point. Indeed, the particle speed at the crest $q(0)$ is
horizontal and is defined as follows:
\begin{equation}\label{eq:q(0)}
\mathit{\Phi}_x\left(0,~\eta(0)\right)=q(0)=c\pm\sqrt{c^2-2\eta(0)},
\end{equation}
$\eta(0)$ being the height of the crest above the still water
level. The ``$-$" sign corresponds to the classical Stokes branch.
The value $\eta_{\max}(0)=c^2/2$ corresponds to the Stokes wave of
limiting amplitude. In this case, the particle speed at the crest
is exactly equal to the wave phase speed: $q_{\max}(0)=c$. Taking
into account both signs in expression (\ref{eq:q(0)}), we assume
that a second branch of solutions should exist apart from the
Stokes waves, at $\eta(0)<\eta_{\max}(0)$. The particle speed at
the crest of a new gravity wave must be greater than $c$ and has
to increase from $c$ to $2c$ while the wave height decreases from
$\eta_{\max}(0)$ to $0$. Moreover, the mean levels of these two
flows relative to the level $y=0$ of still water must also be
different:
\begin{equation} \label{eq:eta0}
\frac{1}{2\pi}
\int_{0}^{2\pi}\eta^{(i)}(x)dx=\eta_{0}^{(i)},~~i=1,~2.
\end{equation}
Thus, the existence of a second branch of solutions of the set of
equations (\ref{eq:Lapl})-(\ref{eq:Inf}) does not contradict
Garabedian's theorem \cite{Garabedian} that gravity waves are
unique if all crests and all troughs are of the same height
because the latter was proved for a flow with the same mean level.

To construct a numerical algorithm we use the method of the
truncated Fourier series and the collocation method, in a plane of
independent spatial variables.

\textit{The method of the Fourier approximations.} Let us
introduce the complex function $R(\theta,~y)$ such that
\begin{equation} \label{eq:R}
\mathit{\Phi}=-i c (R-R^*),~~\mathit{\Psi}= c (R+R^*)
\end{equation}
where $\mathit{\Psi}$ is the stream function, $^*$ is the complex
conjugate. Using the relations
$\mathit\Phi_x=\mathit\Psi_y,~\mathit\Phi_y=-\mathit\Psi_x$, the
kinematic boundary condition (\ref{eq:Kin}) can be presented as
follows:
\begin{equation}\label{eq:KinR}
\frac{d}{dx}
\bigl(R\left(\theta,~\eta\right)+R^*\left(\theta,~\eta\right)-\eta(\theta)\bigr)=0.
\end{equation}

Approximate symmetric stationary solutions of Eq.~(\ref{eq:Lapl},
\ref{eq:Dyn}, \ref{eq:KinR}, \ref{eq:Inf}) are looked for in the
form of the truncated Fourier series with real coefficients
\begin{eqnarray} \label{eq:Ksi}
R(\theta,~y) &=& \sum_{n=1}^N \xi_n \exp \bigl(n (y + i
\theta)\bigr);
\\ \label{eq:Eta}
\eta(\theta)&=&\sum_{n=-M}^M \eta_n \exp(i n \theta),
 ~\eta_{-n}=\eta_n;
\end{eqnarray}
where the Fourier harmonics $\xi_n,~\eta_n$, and the wave speed
$c$ are functions of the wave steepness $A$ determined by the
peak-to-trough height:
\begin{equation} \label{eq:A1}
A=\frac{\eta(0)-\eta(\pi)}{2\pi}=\frac{2}{\pi}\sum_{n=0}^{[M/2]}\eta_{2n+1},
\end{equation}
square brackets designate the integer part. Substitution of
expansions (\ref{eq:Ksi}) and (\ref{eq:Eta}) into the dynamical
and kinematic boundary conditions (\ref{eq:Dyn}), (\ref{eq:KinR})
(the Laplace equation (\ref{eq:Lapl}) and boundary condition
(\ref{eq:Inf}) are satisfied exactly) yields the set of $N+M+1$
non-linear algebraic equations for the harmonics $\xi_n,~\eta_n$,
and the wave speed $c$
\begin{eqnarray} \label{eq:DynF}
\sum_{n_1=1}^N \xi_{n_1} \bigl(f_{n-n_1}^{n_1}+f_{n+n_1}^{n_1}
\bigr)&=&\eta_n,~n=\overline{1,~N};\\ c^2 \sum_{n_1=1}^N n_1
\xi_{n_1} \bigl(f_{n-n_1}^{n_1}+f_{n+n_1}^{n_1}& \nonumber
\\ \label{eq:KinF}
-2\sum_{n_2=1}^N n_2 \xi_{n_2} f_{n+n_1-n_2}^{n_1+n_2}\bigr)
&=&\eta_n,~n=\overline{0,~M};
\end{eqnarray}
where $f_{n}^{n_1}$ are the Fourier harmonics of the exponential
functions $\exp(n_1 \eta(\theta))$:
\begin{equation}\label{eq:f}
f_n^{n_1}=\frac{1}{2\pi}\int_{0}^{2\pi}\exp\bigl(n_1\eta(\theta)-
in\theta\bigr)d\theta,~~f_{-n}^{n_1}=f_n^{n_1}.
\end{equation}
They were being calculated using the fast Fourier transform (FFT).
In addition to these equations, the connection (\ref{eq:A1})
between the harmonics $\eta_n$ and the wave steepness $A$ should
be taken into account.

The set of equations (\ref{eq:DynF}), (\ref{eq:KinF}) was being
solved by Newton's iterations in arbitrary precision computer
arithmetic. Since the non-linearity over $\xi_n$ and $\eta_n$ is
of a different character (polynomial and exponential), the value
of $M$ should be chosen greater than $N$ to achieve good
convergence. A different number of modes for the truncation of the
Fourier series (\ref{eq:Ksi}), (\ref{eq:Eta}) was also used by
Zufiria \cite{Zufiria} in the framework of Hamiltonian formalism.

\textit{The method of collocations.} The harmonics $\xi_n$ of
expansion (\ref{eq:Ksi}) can also be found in another way without
expanding elevation into the Fourier series. In this approach,
Eq.~(\ref{eq:Dyn}) and explicitly integrated Eq.~(\ref{eq:KinR})
are to be satisfied in a number of collocation points
$\theta_j=j\pi/N,~j=\overline{0,~N}$, equally spaced over the half
of one wavelength from the wave crest to the trough, similar to
Rienecker and Fenton \cite{Fenton:col}. This leads to $2N+2$
algebraic equations for the harmonics $\xi_n$, the values of the
elevation $\eta$ at the collocation points, and the wave speed
$c$. To make the numerical scheme better convergent, the greater
number of collocation points may be used in the dynamical boundary
condition (\ref{eq:Dyn}): $M=P N$, $P$ is an integer.

\begin{figure*}
\includegraphics{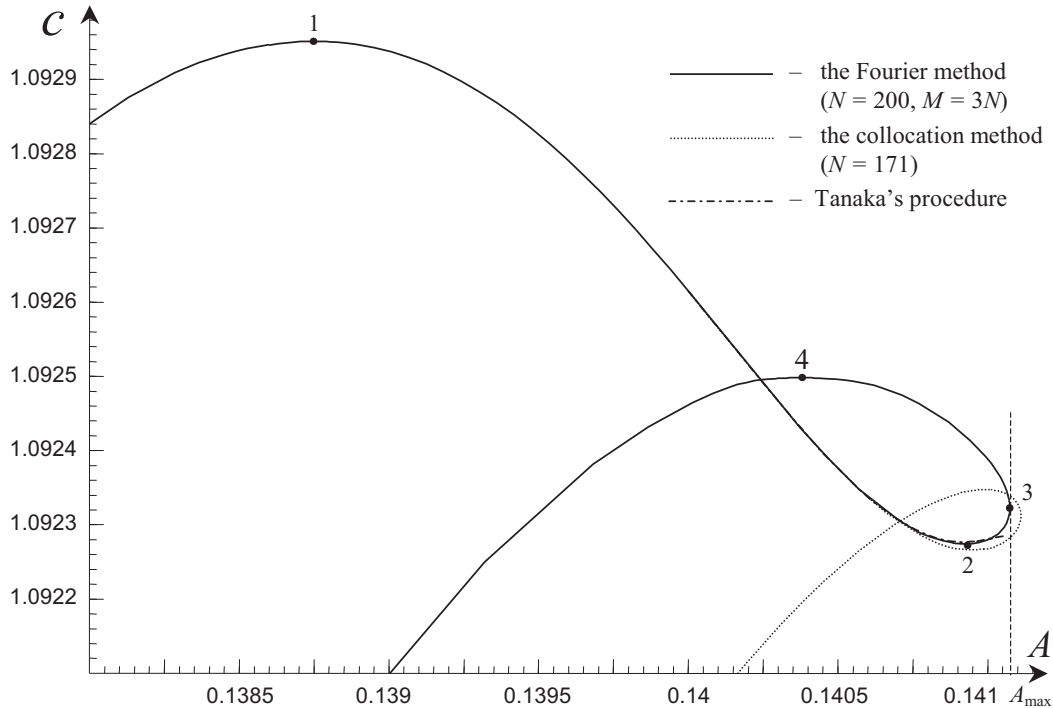}
\caption{\label{fig:c(A)} The wave phase speed $c$ of a surface
wave versus its steepness $A$.}
\end{figure*}

\textit{The results of calculations and discussion.} The
dependence $c(A)$ of the speed of steep gravity waves on their
steepness is shown in Fig.~\ref{fig:c(A)}. Along with the curves
obtained by the Fourier and collocation methods, we included high
accuracy calculations of the Stokes branch by the method of an
inverse plane according to the equations presented in Tanaka's
paper \cite{Tanaka:1983}. In the plot, point 1 ($A=0.13875$) is
the maximum of wave speed, point 2 ($A=0.14092$) is the relative
minimum, point 3 ($A=A_{\max}=0.141074$) corresponds to the
limiting steepness at $N$ and $M$ given. For greater values of $N$
and $M$, $A_{\max} \gtrsim 0.14108$ is obtained. Note, that less
accurate calculations by the collocation method give a greater
value of the limiting steepness which is close to that reported by
Schwartz \cite{Schwartz:1974}.

\begin{table*}
\caption{\label{tab:c(A)} The values of the wave speed $c$ and the
mean water level $\eta_0$ for the steep Stokes and spike waves
with the wave steepness $A$ calculated by the Fourier method. The
values without footnotes were calculated at $N=200,~M=3N$. The
wave speed for the Stokes waves obtained by Tanaka's procedure
\cite{Tanaka:1983} are presented to estimate the accuracy of our
calculations.}
\begin{ruledtabular}
\begin{tabular}{lD{.}{.}{1.10}D{.}{.}{1.10}D{.}{.}{1.7}D{.}{.}{1.10}D{-}{-}{1.3}}
 &\multicolumn{3}{c}{Stokes wave}&\multicolumn{2}{c}{spike wave}\\
\multicolumn{1}{c}{$A$} & \multicolumn{1}{c}{$c$} &
\multicolumn{1}{c}{$c_{~\text{Tanaka}}$} &
\multicolumn{1}{c}{$\eta_0^{(1)} \times 10^7$} &
\multicolumn{1}{c}{$c$}
& \multicolumn{1}{c}{$\eta_0^{(2)} \times 10^7$} \\ \hline \\
0.14 & 1.0926149034 & 1.0926149034 & -2.46\cdot10^{-10} & 1.09246 & -130 \\[3pt]

0.1406 & 1.09233763 & 1.0923377499 & -1.31\cdot10^{-4} & 1.09249 & -60 \\[3pt]

0.14092 & 1.0922742 & 1.0922768392 & -0.0806 & 1.092422 & -22.3 \\
 & 1.0922761\footnotemark[3] &  & -0.0221 & 1.092427\footnotemark[3] & -17.0 \\[3pt]

0.141 & 1.0922796 & 1.0922808596 & -0.385 & 1.092389 & -15.3 \\[3pt]

0.14106 & 1.0922949\footnotemark[1] & 1.0922851047 &
-1.20 & 1.0923548\footnotemark[1] & -8.28 \\
& 1.0922962\footnotemark[2] &  & -1.07 & 1.0923550\footnotemark[2]
& -7.40 \\ [3pt]

0.14107 & 1.0923008\footnotemark[1] & \multicolumn{1}{c}{$-$} &
 -1.64 & 1.0923458\footnotemark[1] & -6.82 \\ [3pt]

0.14108 & 1.0923114\footnotemark[1] & \multicolumn{1}{c}{$-$} &
-2.55 & 1.0923321\footnotemark[1] & -4.89 \\
& 1.0923145\footnotemark[2] &  & -2.45 &
1.0923303\footnotemark[2] & -4.08 \\
\end{tabular}
\end{ruledtabular}
\footnotetext[1]{$N=200,~M=4N$.} \footnotetext[2]{$N=210,~M=4N$.}
\footnotetext[3]{$N=250,~M=3N$.}
\end{table*}

\begin{figure}
\includegraphics{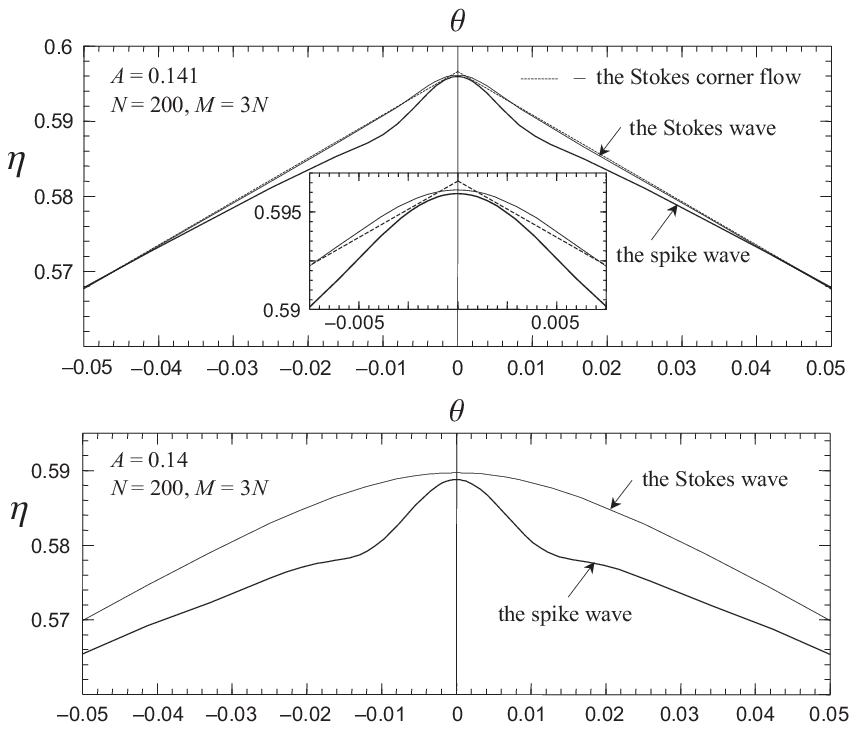}
\caption{\label{fig:profiles} The profiles of a free surface for
the Stokes and spike waves of the same steepness and wavelength
near the crest.}
\end{figure}

The key result of our numerical investigation is that we have
revealed a new branch which arises from the point of the limiting
steepness in the direction of its decreasing so that the loop
2-3-4 is formed. Thus, the point of the limiting steepness seems
to be the point of maximum of $A$, not the breaking point. It
should be noted, that firstly we obtained the new branch by the
Fourier method, and only after that we could track it by the
collocation method using the starting points generated by the
first method. As it is shown in Fig.~\ref{fig:profiles}, the
profile of the new solution near the crest is sharper than the
profile of the Stokes wave of the same steepness. Because of this
we named it ``the spike wave". The difference between the crests
of the Stokes and spike waves becomes stronger as wave steepness
drops relative to the limiting value. In Fig.~\ref{fig:profiles}
the dashed lines designate the exact local Stokes solution (the
Stokes corner flow) which corresponds to the limiting wave with a
maximal value of $A$ (point 3 in Fig.~\ref{fig:c(A)}). In the
immediate vicinity of the crest, the profiles of the almost
highest Stokes and spike waves asymptotically tend to the dashed
lines. This tendency is seen to have the oscillatory character for
both waves. For the Stokes waves such oscillations were
analytically obtained earlier in \cite{Long:al-h1977}.

The values of the wave speed $c$ for the Stokes and spike waves
calculated by the Fourier method at different values of wave
steepness are presented in Table~\ref{tab:c(A)}. The high accuracy
values of the wave speed for the Stokes branch obtained by using
Tanaka's procedure are also included for comparison. One can see,
that for the Stokes branch the accuracy of the Fourier method
gradually decreases as wave steepness increases up to the almost
highest steepness $A=0.14108$. The correspondent value of the wave
speed has only 5 digits stabilized. Note, that Tanaka's procedure
diverges at $A\gtrsim0.141062$. While moving along the new branch
the accuracy becomes still less, and much greater $N$ are needed
to stabilize a greater number of digits. As a result, the form of
the loop in Fig.~\ref{fig:c(A)} has not yet stabilized at $N=200$
and will enlarge with increasing $N$, the cross-section point with
the Stokes branch being moved to the left.

Table~\ref{tab:c(A)} also demonstrates that besides the form near
the crest, the Stokes and spike waves of the same steepness have
different mean water levels $\eta_0$ relative to the still water
level [see Eq.~(\ref{eq:eta0})]. One can see, that at the Stokes
branch $\eta_0^{(1)}$ rapidly descends as $A$ decreases, whereas
$\eta_0^{(2)}$ increases for spike waves. Analysis of dependences
of $\eta_0^{(1)}$, $\eta_0^{(2)}$ on $N$ and $M$ indicates that
they tend to different values at $N,~M\rightarrow\infty$.

At the beginning of the paper we assumed the existence of a new
type of gravity waves for which the speed of a particle at the
crest is greater than wave speed. This property is confirmed by
the calculations presented in Table~\ref{tab:Fx(0)}.
\begin{table}
\caption{\label{tab:Fx(0)} The values of the particle speed at the
crest of the Stokes and spike waves of the same wave steepness
$A$, in a reference frame moving with wave speed. All values were
calculated by the Fourier method at $N=200,~M=4N$.}
\begin{ruledtabular}
\begin{tabular}{lD{.}{.}{1.4}D{.}{.}{1.4}}
&\multicolumn{2}{c}{$q(0)-c$}\\
\multicolumn{1}{c}{$A$} & \multicolumn{1}{c}{Stokes wave}& \multicolumn{1}{c}{spike wave} \\
\hline
0.14092& -0.0370 & 0.0540 \\
0.14106& -0.0121 & 0.0240 \\
0.14107& -0.0076 & 0.0194 \\
0.14108& -0.0003 & 0.0121 \\
\end{tabular}
\end{ruledtabular}
\end{table}

Thus, the spike waves, which we found numerically using two
independent methods, present a new type of gravity waves we looked
for. In the present work, we interested only in the existence of
new stationary solutions and did not investigate their stability.
Profiles of the almost highest Stokes and spike waves differ only
in the vicinity of the crest. This leads us to an assumption that
excitation of spike waves may possibly be connected with the crest
instabilities \cite{Long:al-h1994-2} of the Stokes almost highest
waves. From the other side, sharpening of the crest of a spike
wave, when wave steepness decreases (see
Fig.~{\ref{fig:profiles}}), makes us look for a relation to a
problem of existence of solitary waves on deep water. At present,
all existent experimental observations of surface solitary waves
on deep water are usually interpreted by excitation of internal
waves in stratified ocean \cite{Osborne:scince}. However,
verification of our assumption demands another numerical algorithm
since the ones presented above become ineffective. Finally,
two-valued character of a solution of
Eq.~(\ref{eq:Lapl})-(\ref{eq:Inf}) in the vicinity of the limiting
steepness does not depend on depth, as follows from
Eq.~(\ref{eq:q(0)}). We have recently revealed a second branch for
a layer of finite depth.
\begin{acknowledgments}
We are grateful to Professor D.H. Peregrine for helpful assistance
in calculations of the Stokes waves by Tanaka's procedure and to
Professor C. Kharif for many valuable advices and fruitful
discussions. This research has been supported by INTAS grant
99-1637.
\end{acknowledgments}

\end{document}